\documentclass[conference]{IEEEtran}
\IEEEoverridecommandlockouts

\usepackage{amsmath,amsfonts}
\usepackage[ruled,linesnumbered,lined]{algorithm2e}
\usepackage{algorithmic}
\usepackage{array}
\usepackage[caption=false,font=normalsize,labelfont=sf,textfont=sf]{subfig}
\usepackage{textcomp}
\usepackage{stfloats}
\usepackage{url}
\usepackage{verbatim}
\usepackage{graphicx}
\usepackage{cite}
\usepackage{bm}
\usepackage{bbm}

\ifCLASSINFOpdf
\else
\fi
\begin{document}
%
\title{Transferable Multi-Fidelity Bayesian Optimization for Radio Resource Management}

\author{\IEEEauthorblockN{Yunchuan Zhang, Sangwoo Park, and Osvaldo Simeone \thanks{The work of O. Simeone was supported by European Union’s Horizon Europe project CENTRIC (101096379), by the Open Fellowships of the EPSRC (EP/W024101/1), by the EPSRC project (EP/X011852/1), and by  Project REASON, a UK Government funded project under the Future Open Networks Research Challenge (FONRC) sponsored by the Department of Science Innovation and Technology (DSIT).}}
\IEEEauthorblockA{Centre for Intelligent Information Processing Systems (CIIPS)\\Department of Engineering, King's College London, UK\\
Email: \{yunchuan.zhang, sangwoo.park, osvaldo.simeone\}@kcl.ac.uk}
}


%


\maketitle

\begin{abstract}
Radio resource allocation often calls for the optimization of black-box objective functions whose evaluation is expensive in real-world deployments. Conventional optimization methods apply separately to each new system configuration, causing the number of evaluations to be impractical under constraints on computational resources or timeliness. Toward a remedy for this issue, this paper introduces a multi-fidelity continual optimization framework that hinges on a novel information-theoretic acquisition function. The new strategy probes candidate solutions so as to balance the need to retrieve information about the current optimization task with the goal of acquiring information transferable to future resource allocation tasks, while satisfying a query budget constraint. Experiments on uplink power control in a multi-cell multi-antenna system demonstrate that the proposed method substantially improves the optimization efficiency after processing a sufficiently large number of tasks. 
\end{abstract}



%
\IEEEpeerreviewmaketitle

\section{Introduction}\label{sec: intro}
Radio resource management in modern cellular networks requires the optimization of costly-to-evaluate objectives. For instance, mobile network operators perform time-consuming and expensive experiments to configure the parameters used by policies such as transmission power control \cite{calabrese2018learning}. Machine learning (ML)-assisted optimizers are widely adopted to address these complex problems \cite{simeone2018very,maggi2021bayesian,chen2023neuromorphic}. But conventional ML tools such as Bayesian optimization (BO) may require impractical amounts of real-world data such as channel state information (CSI) samples \cite{zhang2023bayesian}. Moreover, mobility and changes in traffic and channel conditions call for the continual adaptation of resource allocation strategies.

For concreteness, consider the problem of uplink \emph{open loop power control} (OLPC) in a multi-cell multi-antenna communication system \cite{maggi2021bayesian}. In this problem, the optimizer selects a set of power control variables for each new configuration of the system consisting of a different distribution of the users equipments (UEs). When addressed sequentially over time, the resource allocation tasks are statistically correlated, and thus knowledge from one task can be potentially transferred to future tasks \cite{vinyals2016matching}.

BO has been widely applied to various \emph{static} black-box optimization problems in communication systems, such as antenna design \cite{zhou2020trust}, location estimation \cite{jagadeesan2017unifying} and transmission power control \cite{maggi2021bayesian,zhang2023bayesian}. Prior works in the ML literature also addressed BO extensions to \emph{multi-fidelity} \cite{moss2021gibbon} settings, in which evaluations at lower fidelity levels contribute to improving the optimization efficiency as a function of the overall cost budget for evaluating the target objective function. In particular, reference \cite{moss2021gibbon} introduced a general information-theoretic acquisition function enabling a light-weight multi-fidelity BO (MFBO) framework that seeks for the next candidate solution by maximizing the \emph{information gain per unit cost}.

In this paper, we introduce a novel \emph{information-theoretic} acquisition function, namely \emph{multi-fidelity transferable max-value entropy search} (MFT-MES), which balances the need to acquire information about the current task with the goal of collecting information transferable to future tasks. Existing studies that address transferability for BO, including \emph{lifelong BO} \cite{zhang2019lifelong} and \emph{meta-learned BO} \cite{zhang2023bayesian,rothfuss2023meta}, are limited to single-fidelity settings that do not account for the evaluation cost budget. The proposed method is validated on the mentioned application to uplink OLPC in cellular systems with dynamic network topologies.

\begin{figure*}[t]
  \centering
  \centerline{\includegraphics[scale=0.45]{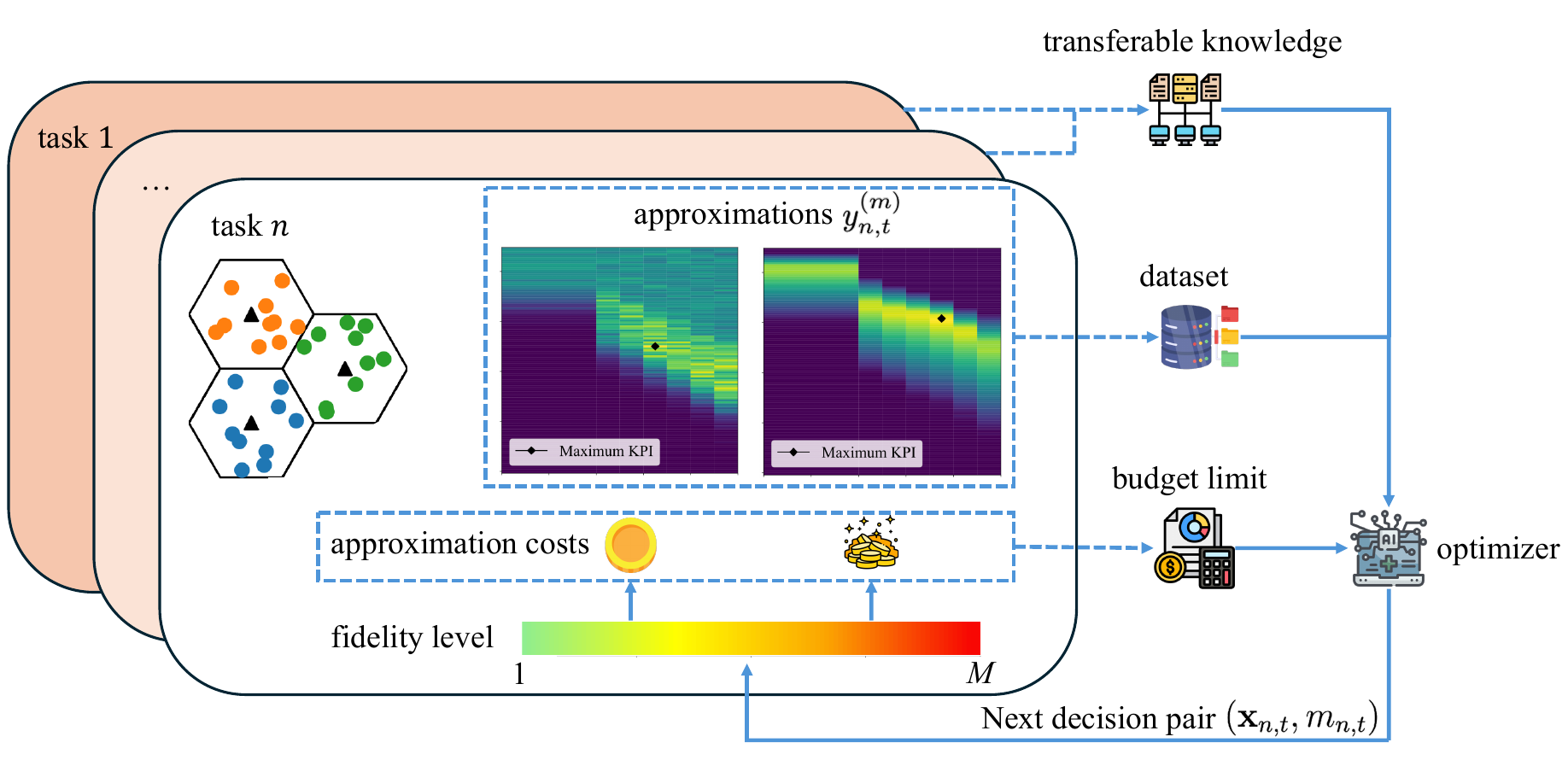}}
  \vspace{-0.2cm}
  \caption{This paper addresses settings with a sequence of black-box optimization tasks using BO with multi-fidelity approximations of the objective function. For the example of interest in this work, each task is specified by a network topology with randomly distributed UEs (colored dots in hexagons). For any current task $n$, over iteration $t=1,2,...,T_n$, the optimizer attempts a solution $\mathbf{x}_{n,t}$ at fidelity level $m_{n,t}$, requiring an evaluation cost $S^{(m)}$. As a result, the optimizer detects noisy feedback $y_{n,t}^{(m)}$ about objective value $f_n^{(m)}(\mathbf{x}_{n,t})$. The goal is to locate the global optimum $\mathbf{x}_n^*$ of objective $f_n(\mathbf{x})$ without exceeding a total simulation cost budget $\Lambda$ for any task $n$.}
  \label{fig: intro flow}
  \vspace{-0.3cm}
\end{figure*}

\section{Problem Formulation}\label{sec: pf}
To exemplify the use of the proposed continual black-box optimization approach, we consider the problem of physical uplink shared channel (PUSCH) transmission power control specified by 3GPP TS 38.213 \cite{3gpp38213}. As in \cite{maggi2021bayesian,zhang2023bayesian}, we focus on a wireless cellular communication system consisting of $N_C$ cells with one base station (BS), each equipped with $N_T$ transmit antennas, and $N_U$ UEs, each with $N_R$ antennas, in each cell. 

At any optimization interval $n$, also referred to as a \emph{task}, the network configuration is defined by a collection of path losses $\{\text{PL}_{n,c,u,c'}\}$ for all UEs $u\in\{1,...,N_U\}$ in cell $c\in\{1,...,N_C\}$ towards the BS in cell $c'\in\{1,...,N_C\}$. Accordingly, for each task $n$, the $N_R\times N_T$ channel matrix between UE $u$ in cell $c$ and the BS in cell $c'$ is modelled as
\begin{align}
    \mathbf{H}_{n,c,u,c'}=10^{\frac{-\mathrm{PL}_{n,c,u,c'}}{20}}\xi_{n,c,u,c'}\mathbf{G}_{n,c,u,c'}, \label{eq: channel matrix}
\end{align}
where the $N_R \times N_T$ fast fading matrix $\mathbf{G}_{n,c,u,c'}$ has i.i.d. standard complex Gaussian entries, and the slow fading coefficient $\xi_{n,c,u,c'}$ depends on whether the transmission is in non-line-of-sight (NLOS) or line-of-sight (LOS) as specified by 3GPP TR 38.901 \cite{3gpp38213}.

For each task $n$, each UE $u$ in cell $c$ calculates its transmit power $P^{\text{TX}}_{n,c,u}$ to the serving BS in the same cell as a function of the OLPC parameters $\mathbf{x}=(P_{0},\alpha)$, whose values are constrained as in Table \ref{table: olpc choices}, as well as of the corresponding path loss $\text{PL}_{n,c,u,c}$ using \cite[Eq. (1)]{zhang2023bayesian}.

For each $n$-th task, with $n=1,2,...$, the objective is to maximize the average sum of the spectral efficiencies (SSE) across all cells, which is expressed as
\begin{align}
    f_n(\mathbf{x})&=\mathbbm{E}_n[\text{SSE}(\mathbf{x},\mathbf{H}_n)]\nonumber\\&=\mathbbm{E}_n\bigg[\sum_{c=1}^{N_C}\sum_{u=1}^{N_U} \log_2\big|\mathbf{I}_{N_R}\nonumber\\&\hspace{0.4cm}+10^{\frac{P^{\mathrm{TX}}_{n,c,u}}{10}}\boldsymbol{\Gamma}_{n,c,u}^{-1}\mathbf{H}_{n,c,u,c}\mathbf{H}_{n,c,u,c}^{\sf H}\big|\bigg],\label{eq: mimo kpi}
\end{align}
where $\mathbbm{E}[\cdot]$ is the expectation of CSI \eqref{eq: channel matrix}, collectively denoted as $\mathbf{H}_n$; $\mathbf{I}_{N_R}$ is the $N_R\times N_R$ identity matrix; and $\boldsymbol{\Gamma}_{n,c,u}$ represents the $N_R\times N_R$ noise-plus-interference covariance matrix for the transmission of UE $u$ towards the serving BS in cell $c$, i.e.,
\vspace{-0.2cm}
\begin{align}
    \boldsymbol{\Gamma}_{n,c,u} &= 10^{\frac{\sigma_z^2}{10}}\mathbf{I}_{N_R}+\sum_{j=1, j\neq u}^{N_U}10^{\frac{P^{\mathrm{TX}}_{n,c,j}}{10}}\mathbf{H}_{n,c,j,c}\mathbf{H}^{\sf H}_{n,c,j,c}\nonumber \\&+\sum_{c'=1,c'\neq c}^{N_C}\sum_{u=1}^{N_U}10^{\frac{P^{\mathrm{TX}}_{n,c',u}}{10}}\mathbf{H}_{n,c,u,c'}\mathbf{H}_{n,c,u,c'}^{\sf H},\label{eq: noise and interference}
\end{align}
with $\sigma_z^2$ as the channel noise power in dB scale. Overall, the problem of interest for any task $n$ can be expressed as the optimization
\begin{align}
    \mathbf{x}_n^*=\arg\max\limits_{\mathbf{x}}f_n(\mathbf{x}).  \label{eq: poptimization target}
\end{align}

The objective function $f_n(\mathbf{x})$ in \eqref{eq: mimo kpi} cannot be directly evaluated, since it requires averaging over the unknown CSI distribution for each task $n$. However, it can be approximated by collecting CSI samples $\{\mathbf{H}_{n,s}\}_{s=1}^{S^{(m)}}$ via the empirical average objective
\begin{align}
     f^{(m)}_{n}(\mathbf{x})=\frac{1}{S^{(m)}}\sum_{s=1}^{S^{(m)}} \text{SSE}(\mathbf{x},\mathbf{H}_{n,s}).
\label{eq: noisy observation}
\end{align}
The evaluation cost depends on the number of samples $S^{(m)}$.

To account for the evaluation cost, we adopt a \emph{multi-fidelity} formulation \cite{le2014recursive}. Specifically, we allow the optimizer to select a fidelity level $m$ within a set of $M$ options, $[1,...,M]$, with each level $m$ corresponding to a  cost $S^{(m)}$.  The costs are ordered from lowest fidelity to highest fidelity as $S^{(1)}\leq\cdots\leq S^{(M)}$.

For each task $n$, the optimizer evaluates the objective $f_n(\mathbf{x})$ during $T_n$ \emph{rounds}, choosing at each round $t=1,...,T_n$ an OLPC parameter $\mathbf{x}_{n,t}$ and a fidelity level $m_{n,t}$. The number of rounds, $T_n$, is dictated by the cost budget $\sum_{t=1}^{T_n}S^{(m_t)}\leq\Lambda$. For each candidate solution $\mathbf{x}_{n,t}$, the optimizer observes a noisy feedback \cite{zhang2023bayesian}
\begin{align}
    y^{(m)}_{n,t} = f_n^{(m)}(\mathbf{x}_{n,t})+\epsilon_{n,t}, \label{eq: noisy obs}
\end{align}
where the observation noise variables $\epsilon_{n,t}\sim\mathcal{N}(0,\sigma^2)$ are independent. The pair $(\mathbf{x}_{n,t+1},m_{n,t+1})$ is chosen by the optimizer based on the past observations
\begin{align}
    \mathcal{D}_{n,t}=\Big\{(\mathbf{x}_{n,1},m_{n,1},y_{n,1}^{(m_1)}),...,(\mathbf{x}_{n,t},m_{n,t},y_{n,t}^{(m_t)})\Big\} \label{eq: current dataset}
\end{align}
for the current task $n$, as well as based on previously collected $n'=1,...,n-1$ datasets $\mathcal{D}_{n-1}=\cup_{n'=1}^{n-1} \mathcal{D}_{n',T_{n'}}$.




\section{Information-Theoretic Multi-Fidelity Bayesian Optimization}\label{sec: mfbo}
In this section, we briefly review the information-theoretic framework introduced in \cite{moss2021gibbon} for a single optimization task.

\subsection{Multi-Fidelity Gaussian Process}\label{ssec: mfgp}
\emph{Multi-fidelity Gaussian process} (MFGP) provides a surrogate model for approximation functions $\{f_n^{(m)}(\mathbf{x})\}_{m=1}^M$ across all fidelity levels \cite{bonilla2007multi}. This is done by placing a GP prior over functions, which is expressed as
\begin{align}
    &(f_n^{(1)}(\mathbf{x}),...,f_n^{(M)}(\mathbf{x}))\sim\mathcal{GP}(0,k_{\boldsymbol{\theta}}((\mathbf{x},m),(\mathbf{x}',m'))), \label{eq: mfgp sample} \\ &\text{with} \quad k_{\boldsymbol{\theta}}\big((\mathbf{x},m),(\mathbf{x}',m')\big)=k_{\boldsymbol{\theta}}(\mathbf{x},\mathbf{x}')\cdot \kappa(m,m'),\label{eq: icm kernel}
\end{align}
where a \emph{radial basis function} (RBF) kernel is adopted for fidelity space kernel $\kappa(m,m')=\exp(-\gamma||m-m'||^2)$; and a parametric kernel $k_{\boldsymbol{\theta}}(\mathbf{x},\mathbf{x}')$ for input space is given by \cite{rothfuss2023meta}
\begin{align}
    k_{\boldsymbol{\theta}}(\mathbf{x},\mathbf{x}')=\exp{(-||\psi_{\boldsymbol{\theta}}(\mathbf{x})-\psi_{\boldsymbol{\theta}}(\mathbf{x}')||_2^2}),\label{eq: sogp para kernel}
\end{align}
in which $\psi_{\boldsymbol{\theta}}(\cdot)$ is a neural network with parameters $\boldsymbol{\theta}$. The notation $k_{\boldsymbol{\theta}}(\mathbf{x},\mathbf{x}')$ makes it clear that the kernel function measures the correlation of function values $f_n(\mathbf{x})$ and $f_n(\mathbf{x}')$.

Let $\mathbf{X}_{n,t}=[\mathbf{x}_{n,1},...,\mathbf{x}_{n,t}]^{\sf T}$ denote the set of inputs queried up to time $t$, and $\mathbf{m}_{n,t}=[m_{n,1},...,m_{n,t}]^{\sf T}$ be the $t\times 1$ vector of queried fidelity levels, receiving $t\times 1$ noisy observation vector $\mathbf{y}_{n,t}=[y^{(m_1)}_{n,1},...,y^{(m_{t})}_{n,t}]^{\sf T}$. Using the $t\times t$ kernel matrix $\mathbf{K}_{\boldsymbol{\theta}}(\mathbf{X}_{n,t},\mathbf{m}_{n,t})$ in which the $(i,j)$-th element $k_{\boldsymbol{\theta}}((\mathbf{x}_{n,i},m_{n,i}),(\mathbf{x}_{n,j},m_{n,j}))$ is defined as in \eqref{eq: icm kernel}, the MFGP posterior mean and variance at any input $\mathbf{x}$ with fidelity $m$ given dataset $\mathcal{D}_{n,t}$ can be expressed as 
\cite{moss2021gibbon}
\begin{subequations}
    \begin{equation}
        \mu^{(m)}_{\boldsymbol{\theta}}(\mathbf{x}|\mathcal{D}_{n,t})=\mathbf{k}_{\boldsymbol{\theta}}(\mathbf{x},m)^{\sf T}(\tilde{\mathbf{K}}_{\boldsymbol{\theta}}(\mathbf{X}_{n,t},\mathbf{m}_{n,t}))^{-1}\mathbf{y}_{n,t},\label{eq: mtgp mean}
    \end{equation}
    \begin{align}
        [\sigma_{\boldsymbol{\theta}}^{(m)}(\mathbf{x}|\mathcal{D}_{n,t})]^2=&k_{\boldsymbol{\theta}}((\mathbf{x},m),(\mathbf{x},m))-\mathbf{k}_{\boldsymbol{\theta}}(\mathbf{x},m)^{\sf T}\nonumber\\&(\tilde{\mathbf{K}}_{\boldsymbol{\theta}}(\mathbf{X}_{n,t},\mathbf{m}_{n,t}))^{-1}\mathbf{k}_{\boldsymbol{\theta}}(\mathbf{x},m),\label{eq: mtgp variance}
    \end{align}\label{eq: mfgp posterior}
\end{subequations}
where $\mathbf{k}_{\boldsymbol{\theta}}(\mathbf{x},m)=[k_{\boldsymbol{\theta}}((\mathbf{x},m),(\mathbf{x}_{n,1},m_{n,1})),\hdots,k_{\boldsymbol{\theta}}((\mathbf{x},m),\\(\mathbf{x}_{n,t},m_{n,t}))]^{\sf T}$ is the $t \times 1$ cross-variance vector; and $\tilde{\mathbf{K}}_{\boldsymbol{\theta}}(\mathbf{X}_{n,t},\mathbf{m}_{n,t})=\mathbf{K}_{\boldsymbol{\theta}}(\mathbf{X}_{n,t},\mathbf{m}_{n,t})+\sigma^2\mathbf{I}_{t}$ is the $t\times t$ Gramian matrix.

\subsection{Multi-Fidelity Max-Value Entropy Search}\label{ssec: mfmes detail}
For brevity of notation, we henceforth omit the dependence on the observation history $\mathcal{D}_{n,t}$ of the MFGP posterior mean $\mu^{(m)}_{\boldsymbol{\theta}}(\mathbf{x}|\mathcal{D}_{n,t})$ and variance $[\sigma_{\boldsymbol{\theta}}^{(m)}(\mathbf{x}|\mathcal{D}_{n,t})]^2$ in \eqref{eq: mfgp posterior}, writing $\mu^{(m)}_{\boldsymbol{\theta}}(\mathbf{x})$ and $[\sigma^{(m)}_{\boldsymbol{\theta}}(\mathbf{x})]^2$, respectively. Throughout this subsection, the parameter vector $\boldsymbol{\theta}$ is fixed. In multi-fidelity max-value entropy search, named GIBBON in \cite{moss2021gibbon}, at each time $t$ the next pair $(\mathbf{x}_{n,t+1},m_{n,t+1})$ is selected so as to maximize the ratio between the \emph{informativeness} of the resulting observation $y_n^{(m)}$ in \eqref{eq: noisy obs} and the cost $S^{(m)}$. 

Informativeness is measured by the \emph{mutual information} between the optimal value $f_n(\mathbf{x}_n^*)=f_n^*$ of the objective and the observation $y_{n}^{(m)}$ at input $\mathbf{x}_{n,t+1}$ with fidelity level $m$. Accordingly, the next pair $(\mathbf{x}_{n,t+1},m_{n,t+1})$ is selected by maximizing the \emph{information gain per unit cost} as
\begin{align}
    (\mathbf{x}_{n,t+1},m_{n,t+1})=\arg\max\limits_{\mathbf{x},m}\frac{I(f_n^*;y_{n}^{(m)}|\mathbf{x},\boldsymbol{\theta},\mathcal{D}_{n,t})}{S^{(m)}}.\label{eq: argmax mfmes}
\end{align}

GIBBON approximates the acquisition function in \eqref{eq: argmax mfmes} as
\begin{align}
    &\alpha_{\boldsymbol{\theta}}(\mathbf{x},m)=\frac{1}{S^{(m)}}\log(\sqrt{2\pi e}\sigma_{\boldsymbol{\theta}}^{(m)}(\mathbf{x}))\nonumber\\&-\frac{1}{|\mathcal{F}|S^{(m)}}\sum_{f_n^*\in\mathcal{F}}\log\Bigg(\sqrt{2\pi e}\sigma_{\boldsymbol{\theta}}^{(m)}(\mathbf{x})\bigg(1-\frac{\phi(\gamma_{\boldsymbol{\theta}}^{(m)}(\mathbf{x},f_n^*))}{\Phi(\gamma_{\boldsymbol{\theta}}^{(m)}(\mathbf{x},f_n^*))}\nonumber\\&\quad\bigg[\gamma_{\boldsymbol{\theta}}^{(m)}(\mathbf{x},f_n^*)+\frac{\phi(\gamma_{\boldsymbol{\theta}}^{(m)}(\mathbf{x},f_n^*))}{\Phi(\gamma_{\boldsymbol{\theta}}^{(m)}(\mathbf{x},f_n^*))}\bigg]\bigg)\Bigg),\label{eq: af closed}
\end{align}
where we have defined
\begin{align}
    \gamma_{\boldsymbol{\theta}}^{(m)}(\mathbf{x},f_n^*)=\frac{f_n^*-\mu_{\boldsymbol{\theta}}^{(m)}(\mathbf{x})}{\sigma_{\boldsymbol{\theta}}^{(m)}(\mathbf{x})};\label{eq: gamma func}
\end{align}
the set $\mathcal{F}=\{f_{n,g}^*\}_{g=1}^G$ collects 
$G$ samples drawn from distribution $p_{\boldsymbol{\theta}}(f_n^*|\mathbf{x},\mathcal{D}_{n,t})$ obtained via Gumbel sampling \cite{wang2017max}; and $\phi(\cdot)$ and $\Phi(\cdot)$ are the probability density function and cumulative density function of a standard Gaussian distribution, respectively.

\begin{table}[t]
\centering
\caption{Allowed values for OLPC parameters}\vspace{-0.2cm}
\label{table: olpc choices}
\begin{tabular}{@{}|c|c|@{}}
\hline
 $P_{0}\; (\mathrm{dBm})$ & $-202, -200, ..., +22, +24$  \\ \hline
 $\alpha$ & $0, 0.4, 0.5, 0.6, 0.7, 0.8, 0.9, 1.0$  \\ \hline
\end{tabular}

\end{table}

\section{Sequential Multi-Task Multi-Fidelity Bayesian Optimization}\label{sec: mft-mes}
As reviewed in the previous section, GIBBON treats each task $n$ separately, which may entail significant CSI measurement costs for new tasks. In this section, we introduce \emph{multi-fidelity transferable max-value entropy search (MFT-MES)} that generalizes GIBBON to account for network topology variations over time modelled as sequential multi-fidelity optimization. 

\subsection{Acquisition Function}\label{ssec: mft aq}
MFT-MES treats the kernel parameter vector $\boldsymbol{\theta}$ as being shared across all tasks. Accordingly, it introduces a term in the GIBBON acquisition function \eqref{eq: argmax mfmes} that promotes the selection of inputs $\mathbf{x}$ and fidelity levels $m$ that maximize the information brought by the corresponding observation $y_{n}^{(m)}$ about the shared parameters $\boldsymbol{\theta}$. The rationale behind this modification is that collecting information about the parameters $\boldsymbol{\theta}$ can potentially improve the optimization process for future tasks $n'>n$.

Accordingly, MFT-MES adopts an acquisition function that measures the mutual information between the observation $y_n^{(m)}$ and not only the optimal value $f_n^*$ for the current task $n$, but also the parameters $\boldsymbol{\theta}$. To this end, unlike GIBBON, this approach views parameters $\boldsymbol{\theta}$ as a random quantity that is jointly distributed with the objective and with the observations. Normalizing by the cost $S^{(m)}$ as in \eqref{eq: argmax mfmes}, MFT-MES specifically addresses the problem
\begin{align}
    (\mathbf{x}_{n,t+1},m_{n,t+1})=\arg\max\limits_{\mathbf{x},m}\frac{I(f_n^*,\boldsymbol{\theta};y_{n}^{(m)}|\mathbf{x},\mathcal{D}_{n-1},\mathcal{D}_{n,t})}{S^{(m)}}\label{eq: smfmes acq}
\end{align}
to select the next query $(\mathbf{x}_{n,t+1},m_{n,t+1})$, where the mutual information is evaluated with respect to the joint distribution
\begin{align}
    &p(f_n^*,f_n^{(m)}(\mathbf{x}),\boldsymbol{\theta},y_n^{(m)}|\mathbf{x},\mathcal{D}_{n-1},\mathcal{D}_{n,t})\nonumber\\&\hspace{-0.1cm}=p(\boldsymbol{\theta}|\mathcal{D}_{n-1},\mathcal{D}_{n,t})p(f_n^*,f_n^{(m)}(\mathbf{x})|\mathbf{x},\boldsymbol{\theta},\mathcal{D}_{n,t})p(y_n^{(m)}|f_n^{(m)}(\mathbf{x})),\label{eq: mft joint dist}
\end{align}
where the distribution $p(\boldsymbol{\theta}|\mathcal{D}_{n-1},\mathcal{D}_{n,t})$ is the posterior distribution of the shared parameters $\boldsymbol{\theta}$. This distribution represents the transferable knowledge extracted from all the available observations at round $t$ for task $n$.

Using the chain rule for mutual information \cite{tm2006elements}, the acquisition function \eqref{eq: smfmes acq} is expressed as
\begin{align}
    &\alpha^{\text{MFT-MES}}(\mathbf{x},m)=\frac{1}{S^{(m)}}[I(\boldsymbol{\theta};y_{n}^{(m)}|\mathbf{x},\mathcal{D}_{n-1},\mathcal{D}_{n,t})\nonumber\\&+\mathbbm{E}_{p(\boldsymbol{\theta}|\mathcal{D}_{n-1},\mathcal{D}_{n,t})}[I(f_n^*;y_{n}^{(m)}|\mathbf{x},\boldsymbol{\theta},\mathcal{D}_{n,t})]].\label{eq: meta acq func}
\end{align}
In \eqref{eq: meta acq func}, the first term is the expected information gain on the global optimum $f_n^*$ that is also included in \eqref{eq: argmax mfmes}, while the second term in \eqref{eq: meta acq func} quantifies transferable knowledge via the information gain about the shared parameters $\boldsymbol{\theta}$. 

\subsection{Evaluating the Acquisition Function}\label{ssec: eval acq func}
Let us write as $H(\mathbf{y}|\mathbf{x})$ for the differential entropy of a variable $\mathbf{y}$ given a variable $\mathbf{x}$. The second term in \eqref{eq: meta acq func} can be written more explicitly as
\begin{align}
    &I(\boldsymbol{\theta};y_n^{(m)}|\mathbf{x},\mathcal{D}_{n-1},\mathcal{D}_{n,t})\hspace{-0.1cm}=H(y_{n}^{(m)}|\mathbf{x},\mathcal{D}_{n-1},\mathcal{D}_{n,t})\nonumber\\&\quad-\mathbbm{E}_{p(\boldsymbol{\theta}|\mathcal{D}_{n-1},\mathcal{D}_{n,t})}[H(y^{(m)}_{n}|\mathbf{x},\boldsymbol{\theta},\mathcal{D}_{n,t})]\label{eq: tk original}\\ &\leq \frac{1}{2}\log(2\pi e \text{Var}(y_{n}^{(m)}))\nonumber\\&\quad-\frac{1}{2}\mathbb{E}_{p(\boldsymbol{\theta}|\mathcal{D}_{n-1},\mathcal{D}_{n,t})}[\log(2\pi e[\sigma_{\boldsymbol{\theta}}^{(m)}(\mathbf{x})]^2)], \label{eq: tk full}
\end{align}
where $\text{Var}(y_{n}^{(m)})$ represents the variance of random variable $y^{(m)}_n$ following the mixture of Gaussian distribution $\mathbb{E}_{p(\boldsymbol{\theta}|\mathcal{D}_{n-1},\mathcal{D}_{n,t})}[ \mathcal{N}(y_n^{(m)}|\mu^{(m)}_{\boldsymbol{\theta}}(\mathbf{x}),\sigma_{\boldsymbol{\theta}}^{(m)2}(\mathbf{x})) ]$; and the first term in \eqref{eq: tk full} bounds the differential entropy of a mixture of Gaussians in \eqref{eq: tk original} via the principle of maximum entropy \cite{moshksar2016arbitrarily}. 

When the distribution $p(\boldsymbol{\theta}|\mathcal{D}_{n-1},\mathcal{D}_{n,t})$ is represented using $V$ particles $\{\boldsymbol{\theta}_1,...,\boldsymbol{\theta}_V\}$, the variance in \eqref{eq: tk full} can be approximated via the asymptotically consistent estimate
\begin{align}
\text{Var}(y_{n}^{(m)})=&\sum_{v=1}^V\frac{1}{V}\big([\sigma^{(m)}_{\boldsymbol{\theta}_v}(\mathbf{x})]^2+[\mu_{\boldsymbol{\theta}_v}^{(m)}(\mathbf{x})]^2\big)\nonumber\\&-(\frac{1}{V}\sum_{v=1}^V\mu_{\boldsymbol{\theta}_v}^{(m)}(\mathbf{x}))^2+\sigma^2.\label{eq: total variance}
\end{align}
Consequently, by plugging \eqref{eq: total variance} into \eqref{eq: tk full}, the second term in \eqref{eq: meta acq func} is replaced by the quantity
\begin{align}
    \Delta\alpha^{\text{MFT-MES}}(\mathbf{x},m)= &\frac{1}{2}\Bigg\{\log\Bigg(2\pi e\bigg(\frac{1}{V}\sum_{v=1}^V\big([\sigma^{(m)}_{\boldsymbol{\theta}_v}(\mathbf{x})]^2\nonumber\\&\hspace{-0.05cm}+[\mu_{\boldsymbol{\theta}_v}^{(m)}(\mathbf{x})]^2\big)-\Big(\frac{1}{V}\sum_{v=1}^V\mu_{\boldsymbol{\theta}_v}^{(m)}(\mathbf{x})\Big)^2\bigg)\Bigg)\nonumber \\ &\hspace{-0.05cm}-\frac{1}{V}\sum_{v=1}^V\log\Big(2\pi e\big[\sigma_{\boldsymbol{\theta}_v}^{(m)}(\mathbf{x})\big]^2\Big)\Bigg\}.\label{eq: upper bound trans knowledge}
\end{align}

Based on \eqref{eq: upper bound trans knowledge}, the proposed MFT-MES selects the next pair $(\mathbf{x}_{n,t+1},m_{n,t+1})$ for the current task $n$ by maximizing the criterion \eqref{eq: af closed} with \eqref{eq: upper bound trans knowledge}, i.e.,
\begin{align}
    (\mathbf{x}_{n,t+1},m_{n,t+1})=&\arg\max\limits_{\mathbf{x},m}\frac{1}{S^{(m)}}\Big[\frac{1}{V}\sum_{v=1}^V\alpha_{\boldsymbol{\theta}_v}(\mathbf{x},m)\nonumber\\&+\beta\Delta\alpha^{\text{MFT-MES}}(\mathbf{x},m)\Big],\label{eq: final af}
\end{align}
where the scaling parameter $\beta\geq0$ assigns the relative weight to the task of knowledge transfer for future tasks.

The tightness of the upper bound in \eqref{eq: tk full} relies on the accuracy of approximated distribution $p(\boldsymbol{\theta}|\mathcal{D}_{n-1},\mathcal{D}_{n,t})$ using particles $\{\boldsymbol{\theta}_v\}_{v=1}^V$. We apply \emph{Stein variational gradient descent} (SVGD) \cite{liu2016stein} to iteratively update each particle $\boldsymbol{\theta}_v^{(r+1)}$ at round $r+1$ as
\begin{align}
    \boldsymbol{\theta}_v^{(r+1)} = \boldsymbol{\theta}_v^{(r)}+\eta\Omega(\boldsymbol{\theta}_v^{(r)}),\label{eq: particle update}
\end{align}
where $\eta$ is the stepsize, and the function $\Omega(\boldsymbol{\theta}_v^{(r)})$ is defined as
\begin{align}
    &\Omega(\boldsymbol{\theta}_v^{(r)})=\frac{1}{V}\sum_{v'=1}^V \Bigl[\nabla_{\boldsymbol{\theta}_{v'}^{(r)}} \tilde{k}(\boldsymbol{\theta}_{v'}^{(r)},\boldsymbol{\theta}_v^{(r)})\nonumber\\&+\tilde{k}(\boldsymbol{\theta}_{v'}^{(r)},\boldsymbol{\theta}_v^{(r)})\nabla_{\boldsymbol{\theta}_{v'}^{(r)}} \Big(\ell(\boldsymbol{\theta}_{v'}^{(r)}|\mathcal{D}_{n,t})+\log p_n(\boldsymbol{\theta}_{v'}^{(r)})\Big)\Bigr],\label{eq: smooth func}
\end{align}
with the \emph{negative marginal log-likelihood} of parameters $\boldsymbol{\theta}$ obtained as
\begin{align}
    \ell(\boldsymbol{\theta}|\mathcal{D}_{n,t})&=-\frac{1}{2}(t\log2\pi-\log|\tilde{\mathbf{K}}_{\boldsymbol{\theta}}(\mathbf{X}_{n,t},\mathbf{m}_{n,t})|\nonumber\\&\quad\,-\mathbf{y}_{n,t}^{\sf T}(\tilde{\mathbf{K}}_{\boldsymbol{\theta}}(\mathbf{X}_{n,t},\mathbf{m}_{n,t}))^{-1}\mathbf{y}_{n,t}).\label{eq: mtgp mml}
\end{align}
The prior $p_n(\boldsymbol{\theta})$ for the current task $n$ is obtained based on the kernel density estimation (KDE) of the posterior distribution $p(\boldsymbol{\theta}|\mathcal{D}_{n-1})$ obtained by using particles $\{\boldsymbol{\theta}_1,...,\boldsymbol{\theta}_V\}$ produced at the end of the previous task $n-1$ \cite{bishop2006pattern}. Further details can be found in the journal version \cite{zhang2024multi}.

\section{Numerical Results}\label{sec: exp}
In this section, we empirically evaluate the performance of the proposed MFT-MES on the OLPC tasks introduced in Sec. \ref{sec: pf}. Throughout this section, the cellular communication system has $N_C=3$ cells, $N_U=10$ UEs with $N_T=4$ transmit antennas each, and $N_R=16$ receiving antennas at BS. The collections of path loss $\{\text{PL}_{n,c,u,c'}\}$ and channel fading are simulated following the urban microcellular (UMi) street canyon scenario specified in 3GPP TR 38.901 \cite{3gpp38901}.

\begin{figure}[t]
  \centering
  \centerline{\includegraphics[scale=0.33]{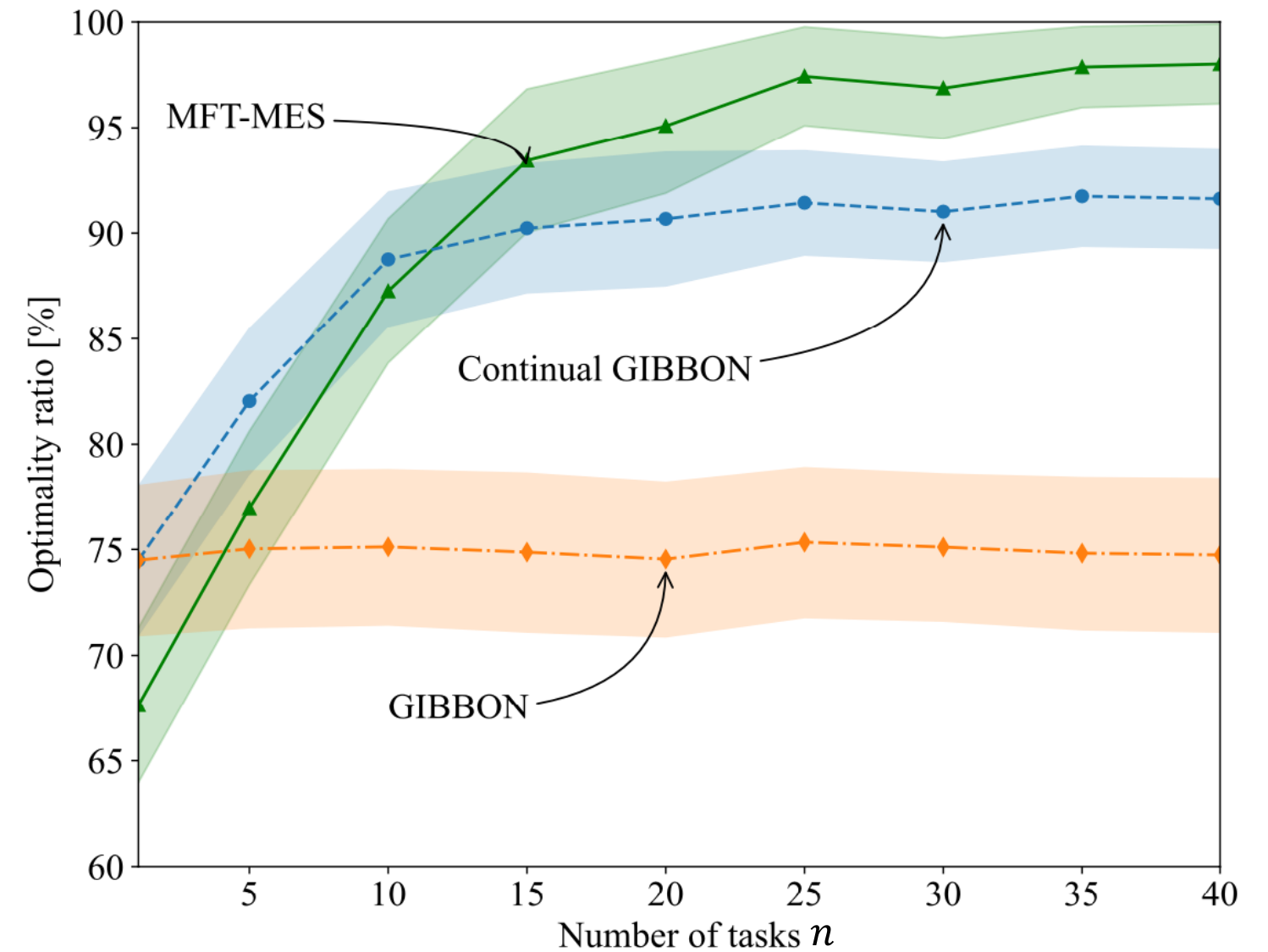}}
  \vspace{-0.3cm}
  \caption{Optimality ratio \eqref{eq: optimality ratio} against the number of tasks, $n$, for GIBBON, Continual GIBBON, and MFT-MES ($\beta=1.6$) with $V=10$ particles.}\label{fig: rrm sr vs task}
\vspace{-0.5cm}
\end{figure}

For the configuration of network topology, in each OLPC task, we set the coverage radius of each serving BS to be 200 meters and uniformly generate random positions of UEs in the interval $[18,200]$ meters; the cost model is set as $S^{(1)}=10,S^{(2)}=20,S^{(3)}=50$, and $S^{(4)}=100$; and total query cost budget is set to $\Lambda=2000$.

The benchmarks considered in all experiments are GIBBON \cite{moss2021gibbon} and \emph{Continual GIBBON}, which adopts $\beta=0$ in \eqref{eq: final af}, thus not extracting transferable knowledge among tasks. Unlike GIBBON \cite{moss2021gibbon}, Continual GIBBON share with MFT-MES the use of SVGD update \eqref{eq: particle update}, but particles $\{\boldsymbol{\theta}_v\}_{v=1}^V$ are carried from one task to the next, in a continual learning manner \cite{zhang2024multi}.

We initialize the MFGP model \eqref{eq: mfgp posterior} with 10 random observations across all fidelity levels, and the observation noise variance is $\sigma^2=0.83$. Following \cite{zhang2023bayesian}, the performance of each method is measured by the \emph{optimality ratio}
\begin{align}
    \max\limits_{t=1,..,T_n}\frac{f^{(M)}_n(\mathbf{x}_{n,t})}{f_n^*},\label{eq: optimality ratio}
\end{align}
which evaluates the best function of the optimal value $f_n^*$ attained during the optimization process. Finally, we set $V=10$ particles for Continual GIBBON and MFT-MES. All results are averaged over 50 random realizations of observation noise signals and $n$ OLPC tasks in the sequence, with error bars to encompass $90\%$ confidence level.

\begin{figure}[t]
  \centering
  \centerline{\includegraphics[scale=0.3]{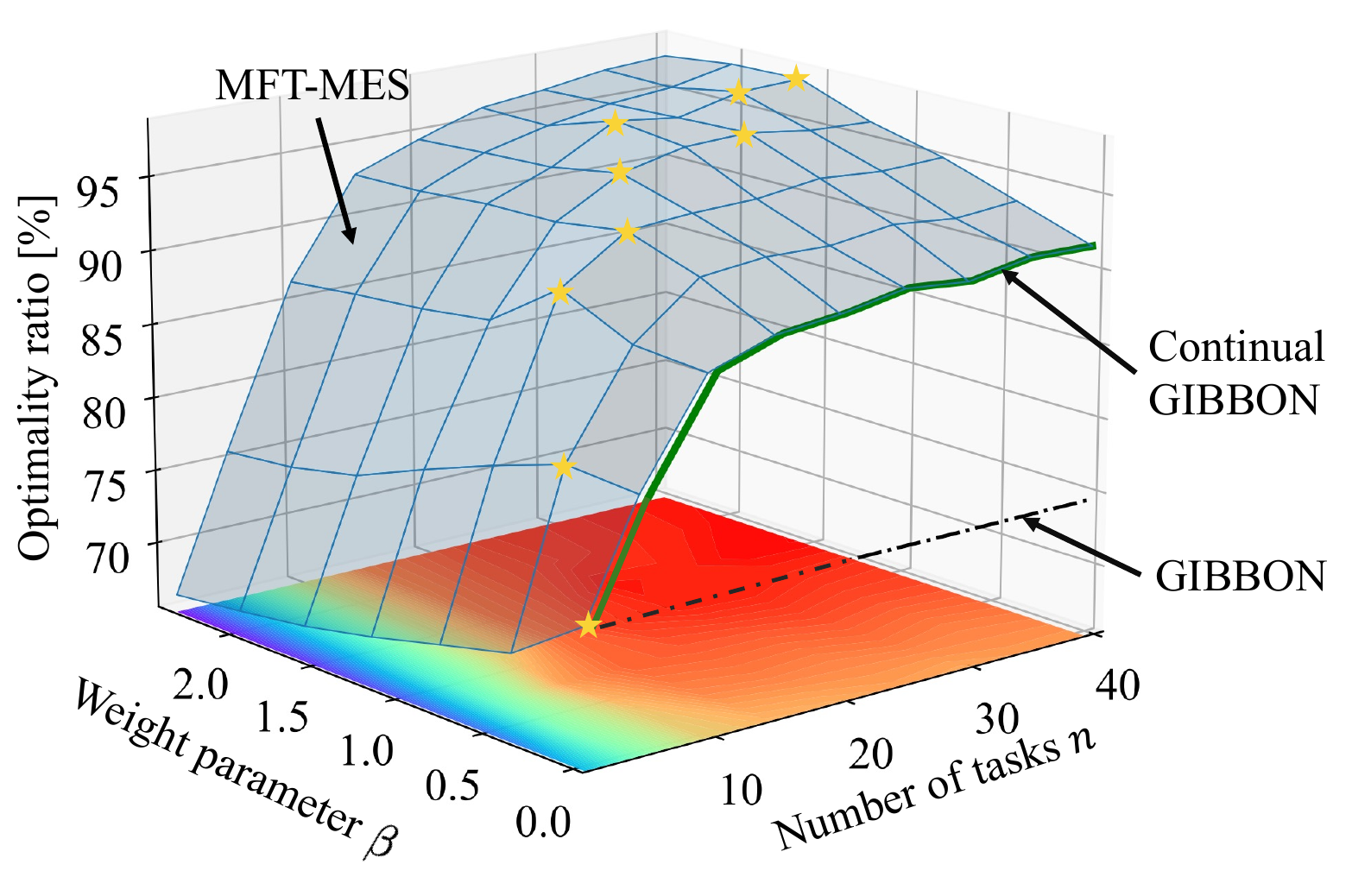}}
  \vspace{-0.2cm}
  \caption{Optimality ratio against weight parameter $\beta$ and the number of tasks, $n$, for GIBBON (black dash-dotted line), Continual GIBBON (green solid line), and MFT-MES (blue surface). The optimal values of $\beta$ at the corresponding number of tasks $n$ are labeled as gold stars. The number of particles for Continual GIBBON and MFT-MES is set to $V=10$.}\label{fig: impact beta}

\end{figure}

In Fig. \ref{fig: rrm sr vs task}, we set the weight parameter $\beta=1.6$ for MFT-MES and plot the optimality ratio \eqref{eq: optimality ratio} as a function of the number of tasks, $n$.  The performance of GIBBON is limited by the lack of information transfer across tasks. In contrast, the performance of both Continual GIBBON and MFT-MES benefits from information transfer. Furthermore, MFT-MES outperforms Continual GIBBON after processing 12 tasks. Through a judicious choice of input and fidelity level targeting the shared parameters $\boldsymbol{\theta}$, at the end of 40-th task, MFT-MES provides, approximately, a $7\%$ gain in terms of optimality ratio over Continual GIBBON, and a $23\%$ gain over GIBBON.

The impact of the weight parameter $\beta$ on the optimality ratio as a function of the number of tasks is studied in Fig. \ref{fig: impact beta}. Recall that the performance of GIBBON, as well as of Continual GIBBON, do not rely on the weight parameter $\beta$, which is an internal parameter of the MFT-MES acquisition function \eqref{eq: final af}. The optimal value of $\beta$, which maximizes the optimality ratio, is marked with a star. 

It is observed that it is generally preferable to increase the value of $\beta$ as the number of tasks $n$ grows larger. This is because a larger $\beta$ favors the selection of inputs and fidelity levels that focus on the performance of future tasks, and this provident approach is more beneficial for longer time horizons. For example, when the sequence of tasks is short, i.e., when $n<5$, the choice $\beta=0$, i.e., Continual GIBBON attains best performance; while MFT-MES with weight parameter $\beta=1.6$ produces best performance given $n=25$ tasks in the sequence.

\bibliographystyle{IEEEtran}

\bibliography{refer}

\end{document}